\documentclass{PoS}
\usepackage{multicol}
\usepackage{multirow}

\newcommand{\bea}{\begin{eqnarray}}
\newcommand{\eea}{\end{eqnarray}}
\newcommand{\be}{\begin{equation}}
\newcommand{\ee}{\end{equation}}

\newcommand{\mlog}{{\rm log}}

\newcommand{\ce}{\chi_\eta}
\newcommand{\ci}{\chi_l}

\newcommand{\ciii}{\chi_s}

\newcommand{\Sig}{\Sigma}

\newcommand{\tr}{{\rm Tr}}
\newcommand{\mres}{m_{\rm res}}

\newcommand{\Dslash}{ \rlap{/}\kern-2.0pt D}

\title{Chiral extrapolations in 2+1 flavor domain wall fermion simulations}

\ShortTitle{Chiral extrapolations in 2+1 flavor domain wall fermion simulations}

\author{RBC and UKQCD Collaborations:  
	\speaker{Meifeng Lin}$^a$\\
	\llap{$^a$} Department of Physics, Columbia University, New York, NY 10027, USA \\
	E-mail: \email{mflin@phys.columbia.edu}
	}
\abstract{%
Simulations with 2+1 flavors of domain wall fermions provide us with
the opportunity to compare the lattice data directly to the predictions of
continuum chiral perturbation theory, up to corrections from the residual
chiral symmetry breaking, $m_{\rm res}$, and ${\cal O}(a)$ lattice
artefacts, which are relatively small for domain wall
fermions.  We present preliminary results for the pseudoscalar meson
masses  and decay constants from partially quenched simulations and
examine the next-to-leading order chiral extrapolations at small quark
masses.  The simulations were carried out on two lattice volumes :  $16^3\times 32$ and $24^3\times 64$, with the lattice spacing fixed at about 0.1 fm. The subtleties of the chiral fits are discussed. We also explore the roles of $m_{\rm res}$ and ${\cal O}(a)$ terms in the NLO chiral expansions and their effects on the chiral extrapolations for the pseudoscalar masses and decay constants.
}

\FullConference{XXIVth International Symposium on Lattice Field Theory\\
                July 23-28, 2006\\
                Tucson, Arizona, USA}

\begin{document}


\label{sec:introduction}
Numerical simulations in lattice QCD are essential for understanding
the non-perturbative aspects of the Standard Model. However, our
ability to obtain the values for physical observables is constrained
by several factors. One among those is that the input quark masses in
the simulations are far heavier than the physical values of the up and
down quarks, due to the limitations of present computing powers. To
deal with this requires extrapolations from the simulation results to
the physical points. Chiral perturbation theory~\cite{Gasser:1984gg}
(ChPT) provides a solid theoretical foundation for these extrapolations. 

However, uncertainty remains concerning the range of quark masses
ChPT formulas can be applied to the numerical data, partly because the
quark masses in the simulations were too heavy until recently. With
the emergence of powerful computers, progress has been made towards
the answer. Simulations with both Wilson fermions~\cite{Farchioni:2004tv} and Staggered fermions~\cite{Aubin:2004fs} have shown consistency between the numerical data and ChPT. Both fermion formulations have limitations, however. Wilson fermions have large chiral symmetry breaking effects at finite lattice spacing, and many new parameters need to be introduced to account for these effects. Staggered fermions even break flavor symmetry, and they require the application of staggered chiral perturbation theory to guide the extrapolations. (For a recent review, see Ref.~\cite{Bar:2004xp}.) In both cases the extra parameters needed complicate the fits and may obscure some systematic effects. 

The domain wall fermion (DWF) and the overlap fermion formulations provide the
cleanest way to probe the chiral limit for various quantities. They both
preserve flavor symmetry at finite lattice spacing, and the breaking of chiral
symmetry for domain wall fermions is exponentially suppressed by
$L_s$, the size of the extent in the fifth dimension. The amount of
the residual chiral symmetry breaking, denoted as $m_{\rm res}$, can be easily measured in the numerical simulations. In this proceedings we will show some progress in the chiral extrapolations for the pseudoscalar meson masses and decay constants from a series of 2+1 flavor domain wall fermion simulations.

\vspace{-0.2cm}
\section{Chiral Expansions for Domain Wall Fermions}
\label{sec:DWF}
 To see how the chiral extrapolations for domain wall fermions should be taken, we need to understand how $m_{\rm res}$ enters the chiral expansions. The axial Ward identity for DWF~\cite{Furman:1994ky} is
\begin{equation}
  \Delta_\mu \langle {\cal A}^a_\mu(x) O(y) \rangle = 
  2m_f \langle J^a_5(x) O(y) \rangle + 2 \langle {J^a_{5q}(x)} O(y)
  \rangle + i \langle \delta^a O(y) \rangle.
  \label{eq:AWI}
\end{equation}
where ${\cal A}^a_\mu(x)$ is the (partially) conserved axial
current. The term containing $J^a_{5q}$ involves fields in the 5th dimension, and
measures the overlap between the quark states with opposite
chiralities that are bound to the two walls at the boundaries of the 5th dimension. As the lattice spacing $a$ goes to 0, we expect to
recover the continuum theory. Thus for low energy Green's functions, $J^a_{5q}$ must be related to the pseudoscalar density $J^a_5$ by
\be
 J^a_{5q} = \mres J^a_5 + {\cal O}(a). 
\ee
With this identity, the terms associated with the input quark mass $m_f$ and the residual mass $\mres$ are indistinguishable. Only the ``effective'' quark mass $m_f + \mres$ plays a role. 
We can then write down the Symanzik effective action to ${\cal O}(a)$ for domain wall fermions:

\begin{equation}
  S_{eff} = \int d^4x \lbrack \bar{\psi}(x) ( i\ \Dslash - {m_q}) \psi(x) \rbrack + {a e^{-\alpha L_s}} c_{dwf} \bar{\psi}(x)\sigma^{\mu\nu}F_{\mu\nu}\psi(x)
\label{eq:Symanzik}
\end{equation}
where $m_q = m_f + \mres$. This is quite similar to Wilson fermions~\cite{Rupak:2002sm}, but the coefficient of the dimension-5 Pauli term is exponentially suppressed by $L_s$ for domain wall fermions. 

To construct the chiral expansions as done in~\cite{Rupak:2002sm}, we need to do the power counting properly. By dimensional analysis, a rough estimate for the size of the Pauli term is 
 
\begin{equation}
ae^{-\alpha L_s} c_{dwf} \sim \mres (a\Lambda_{QCD})^2 \sim {0.1\mres},
\end{equation}
assuming $a$ is about 0.1 fm and $\Lambda_{QCD}\approx 500$ MeV. Thus as long as $\mres$ is not extremely large,
which it is not in our simulations, the Pauli term is a higher order effect compared to the mass contribution. To leading order, only those terms of ${\cal O}(m_q)$ appear in the \emph{chiral Lagrangian}. However, to next-to-leading order, both terms of ${\cal O}(m_q^2)$ and ${\cal O}(ae^{-\alpha L_s})$ contribute, and the chiral Lagrangian becomes
\begin{equation}
  {\cal L}_{NLO}^{a} = 
  {{f^2 \over 8} \tr\lbrack \Sig {\rho} + (\Sig {\rho})^\dagger \rbrack} + {\cal L}_{NLO}^{0} 
  \label{eq:Lagrangian}
\end{equation}
where the superscript $a$ indicates the expression at finite lattice spacing and $0$ stands for the form in the continuum limit. $\rho$ is a spurion field coming from the Pauli term in Eq.~\ref{eq:Symanzik} and has the same chiral transformation as the ordinary mass field. This corrects the continuum chiral formulas for the pseudoscalar masses and decay constants in the following way:
\begin{itemize}
\item The pseudoscalar masses pick up an additional constant term which is proportional to $a e^{-\alpha L_s}$;
\item The expression for the decay constants does not change.
\end{itemize}
In the case of 2+1 flavor sea quarks and degenerate valence quarks, the final expressions, at the chiral scale $\mu$, can be written as
\begin{eqnarray}
  {M_{PS}^2}&=& {\rho} + {\chi_V \Big \lbrace 1+ \frac{48}{{f^2}}({2 L_6-L_4}) \bar\chi
    +\frac{16}{{f^2}}({2L_8-L_5}) \chi_V }\nonumber \\
  & &{+\frac{1}{24{f^2}\pi^2 } \big \lbrack
    \frac{2\chi_V-\ci-\ciii}{\chi_V-\ce}\chi_V\mlog\frac{\chi_V}{\mu} } { - \frac{(\chi_V-\ci)(\chi_V-\ciii)}{(\chi_V-\ce)^2}\chi_V\mlog\frac{\chi_V}{\mu}}
  \nonumber \\
  & &{ +\frac{(\chi_V-\ci)(\chi_V-\ciii)}{\chi_V-\ce}(1+\mlog\frac{\chi_V}{\mu})} {+\frac{(\ce-\ci)(\ce-\ciii)}{(\chi_V-\ce)^2}\ce\mlog\frac{\ce}{\mu}
    \big \rbrack \Big \rbrace }
  \label{eq:mpi_NLO}
\end{eqnarray}
\begin{eqnarray}
  f_{PS}& = & {f} \Big \lbrace 1 + \frac{8}{{f^2}} (3 {L_4} \bar{\chi} + {L_5} \chi_V) \nonumber \\
  & &- \frac{1}{16 \pi^2 {f^2}} \Big\lbrack {(\chi_V + \ci)}\mlog\frac{\chi_V+\ci}{2\mu} +
  \frac{\chi_V + \ciii}{2} \mlog\frac{\chi_V+\ciii}{2\mu}\Big\rbrack \Big \rbrace.
  \label{eq:fpi_NLO}
\end{eqnarray}
The subscripts $l$ and $s$ denote the light and strange sea quarks, and $V$ denotes the valence quarks. $\chi_i = 2 B_0 m_i$, for $ i = V,\ l,\ s,\ {\rm or} \ \eta,\ 
\ce = \frac{1}{3}(\ci+2\ciii)$ and $\bar{\chi} =
\frac{1}{3}(2\ci+\ciii)$. Note that here $m_i$ is the sum of the input quark mass and $\mres$, as implied by Eq.~\ref{eq:Symanzik}.

\section{Numerical Results}
Preliminary work on the chiral extrapolations for domain wall fermions
has been reported in ~\cite{Lin:2005gh}. There we suffered from low
statistics and a large residual mass, thus no reliable NLO chiral fits
could be produced. In the past year, the RBC and UKQCD collaborations
have generated large ensembles of 2+1 flavor domain wall fermion
lattices with 5-dimensional volumes of $16^3\times32\times16$ and
$24^3\times64\times16$. These ensembles were generated using the exact
RHMC algorithm~\cite{Clark:2004cp}, and the Iwasaki gauge action 
with $\beta = 2.13$ was used. For details of the generation of these
ensembles and their properties, see Ref.~\cite{Mawhinney:lat06}. The
lengths of the ensembles are typically 4000 MD time units.  The
inverse lattice spacing is about 1.6 GeV~\cite{Tweedie:lat06}~\cite{Limin:lat06}, which makes the the spatial sizes of the lattices 2 fm and 3 fm. With the quark masses in our simulations, the finite size effects in these volumes are small compared to the statistical errors. Thus in the following discussions, they are neglected.

\subsection{Measurement Details}
Various source smearing techniques have been used to do measurements on these ensembles~\cite{Tweedie:lat06}. The results presented here are all obtained from the Coulomb gauge fixed source with a spatial extent of $16^3$. Partially quenched measurements with valence masses from 0.01 to 0.04 were performed for both volumes. An additional lighter mass point of 0.005 was included for the $24^3\times64$ lattices . The number of measurements for each ensemble is shown in Table~\ref{tab:meas}. 


\begin{table}[ht]
\centering
\begin{tabular}{c|c|cc}
\hline
\hline
\multirow{2}{*}{$m_l/m_s$} & \multirow{2}{*}{$am_{val}$} & \multicolumn{2}{c}{\# measurements [sources]} \\

 & & $16^3\times32$ & $24^3\times64$ \\
\hline
0.01/0.04 & (0.005) 0.01 0.02 0.03 0.04 &  704 [2] & 60 [2]  \\
0.02/0.04 & (0.005) 0.01 0.02 0.03 0.04 & 710 [2] & 53 [2] \\
0.03/0.04 & (0.005) 0.01 0.02 0.03 0.04 & 704 [2] & 55 [2] \\
\hline
\hline
\end{tabular}
\caption{Parameters for the measurements.}
\label{tab:meas}
\end{table}

Note that the normalization of the Coulomb gauge fixed wall source is not known analytically. In contrast to the conventional method of determining $f_{PS}$ using a point-like source~\cite{Blum:2000kn}, we need to compute the relative normalization of the source operator to the conserved current. Bearing this in mind, we can extract the decay constants by 
\begin{equation}
  f_{PS}^2 = 2Z_A^L Z_A^W \frac{ \langle 0 | A_0^L | PS \rangle \langle
    PS | A_0^W | 0 \rangle}{M_{PS}}
\end{equation}
where the superscript W(L) denotes the wall(local) operators. And 
\begin{equation}
  Z_A^L  =  \frac{\sum_{\vec{x}}\langle {\cal A}_0 (\vec{x},t) J_5^W(0) \rangle }{\sum_{\vec{x}}\langle  A_0^L (\vec{x},t) J_5^W(0) \rangle}, 
  Z_A^W  =  \frac{\sum_{\vec{x}}\langle {\cal A}_0 (\vec{x},t) J_5^W(0) \rangle }{\sum_{\vec{x}}\langle  A_0^W (\vec{x},t) J_5^W(0) \rangle}.
\end{equation}
This method gives statistically more accurate results than the point-like sources. The statistical errors for $M_{PS}$ and $f_{PS}$ measured in these ensembles are typically below 1\%. 
\subsection{Chiral Fits at NLO}
To perform the chiral extrapolations using Eq.~\ref{eq:mpi_NLO} and ~\ref{eq:fpi_NLO}, we need to know the value of $\mres$ in the first place. The ratios of the ``mid-point'' correlator to the pseudoscalar correlator are computed on the lattice for each quark mass. And $\mres$ is defined as the value in the chiral limit. With $L_s = 16$, $\mres$ is about 0.003 for these ensembles~\cite{Tweedie:lat06}. We consider two cases for the chiral fits: (I) the ${\cal O}(a)$ corrections are neglected, \textit{i.e.}, the continuum chiral formulas are applied; (II) the full formulas of Eq.~\ref{eq:mpi_NLO} and Eq.~\ref{eq:fpi_NLO} are used. They are discussed in turn as follows. 

\subsubsection{Fits to continuum chiral formulas}
\begin{figure}[ht]
\begin{center}
\begin{tabular}{lr}
\includegraphics*[angle=-90,width=0.45\textwidth,clip]{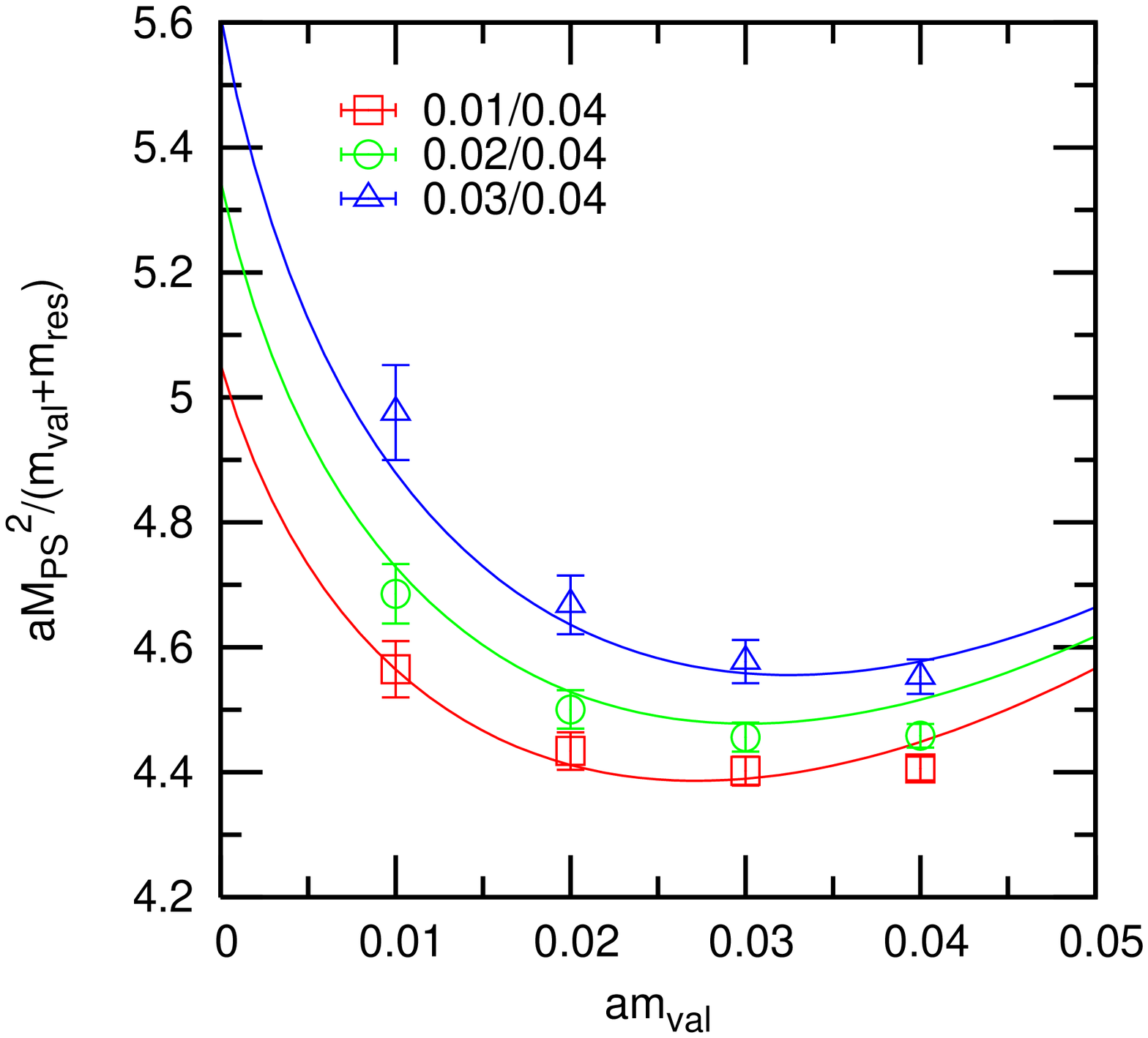} &
\includegraphics*[angle=-90,width=0.45\textwidth,clip]{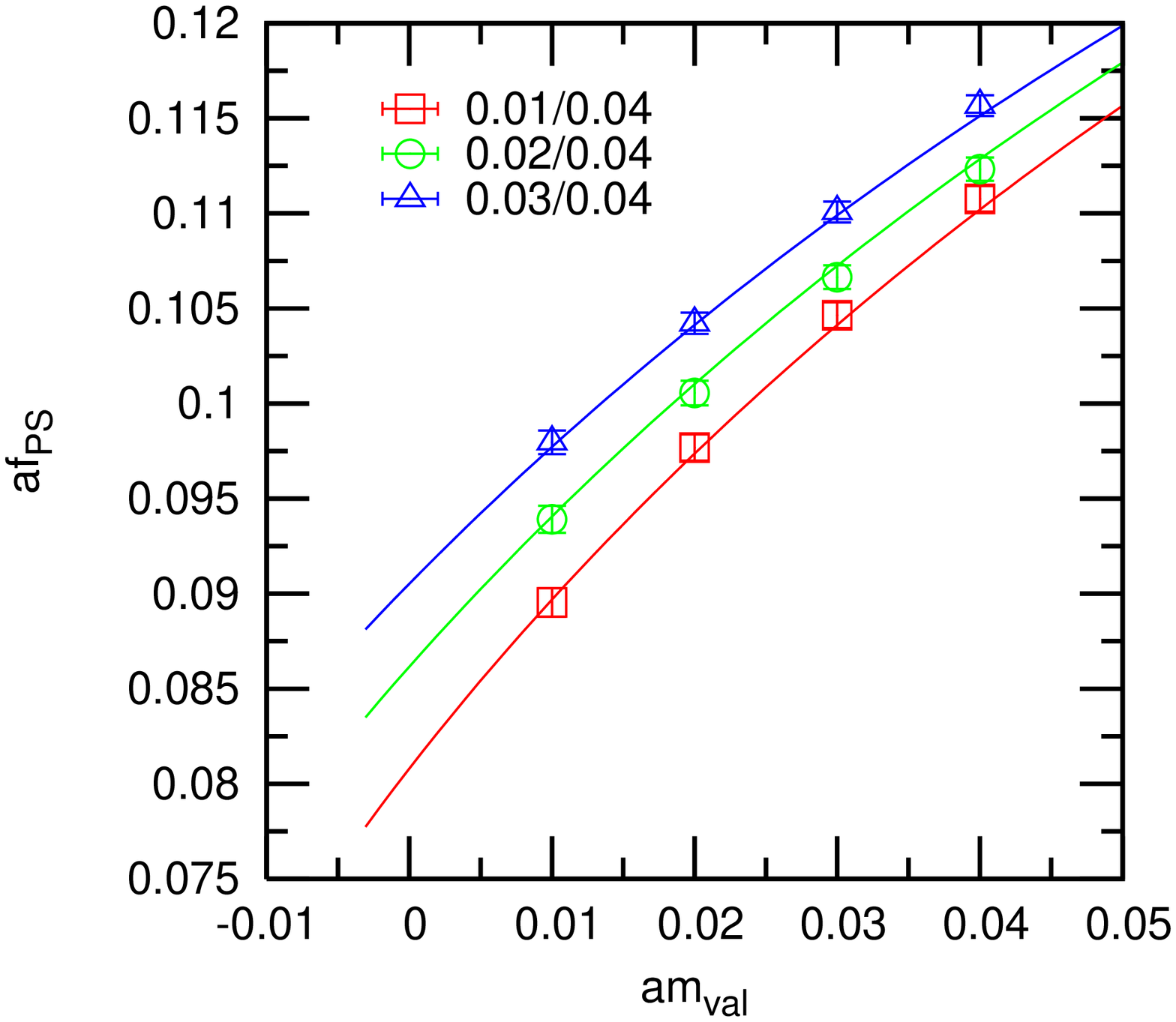}
\end{tabular}
\caption{Partially quenched NLO chiral fits on the $16^3\times32$ ensembles. The fits to $M_{PS}^2$ and $f_{PS}$ are carried out \emph{independent of each other}. The data points at $am_{val} = 0.04$ are excluded from the fits.}
\label{fig:independent_fits}
\end{center}
\end{figure}

\begin{figure}[ht]
\begin{center}
\begin{tabular}{cc}
\includegraphics*[angle=-90,width=0.45\textwidth,clip]{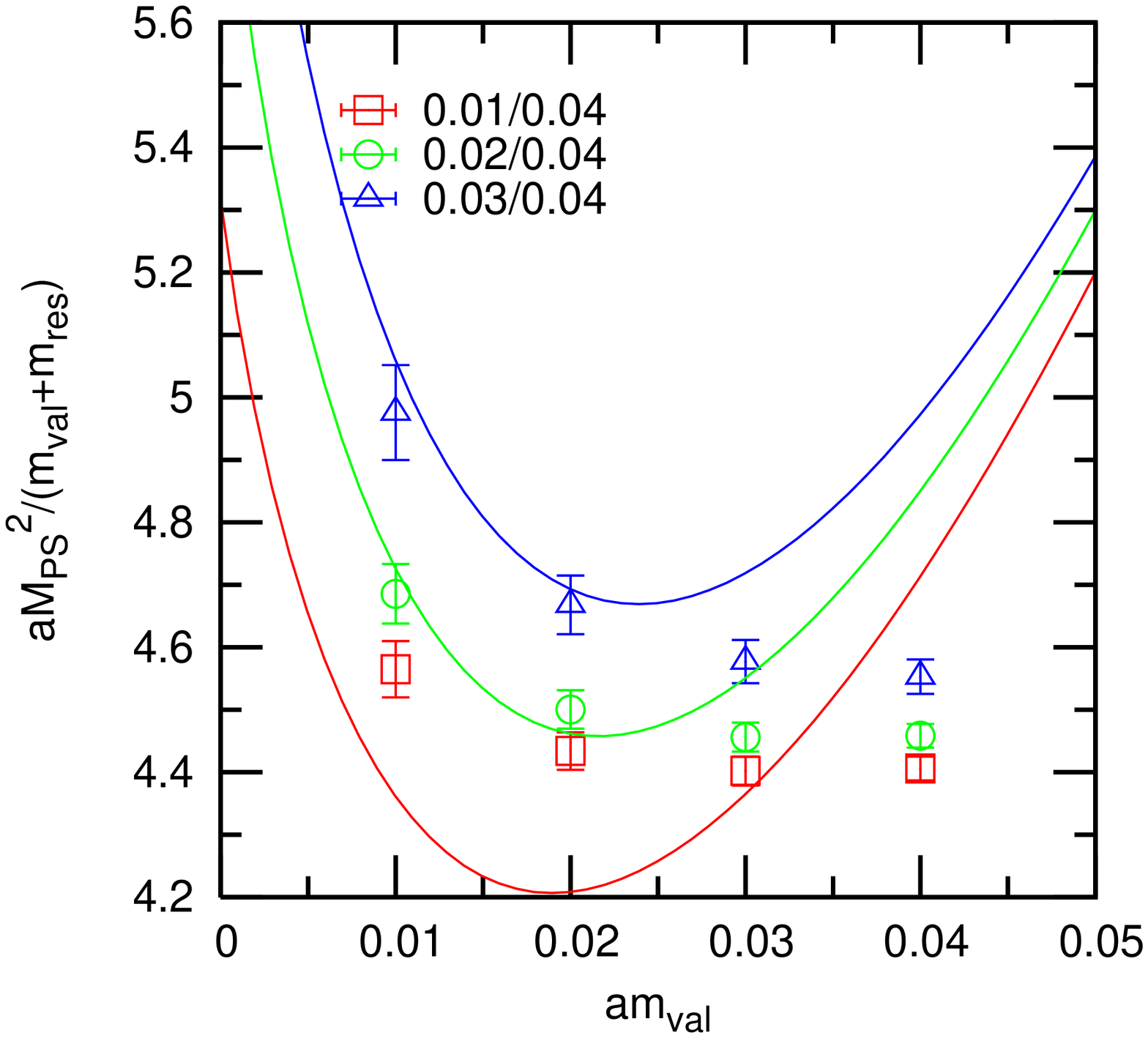} &
\includegraphics*[angle=-90,width=0.45\textwidth,clip]{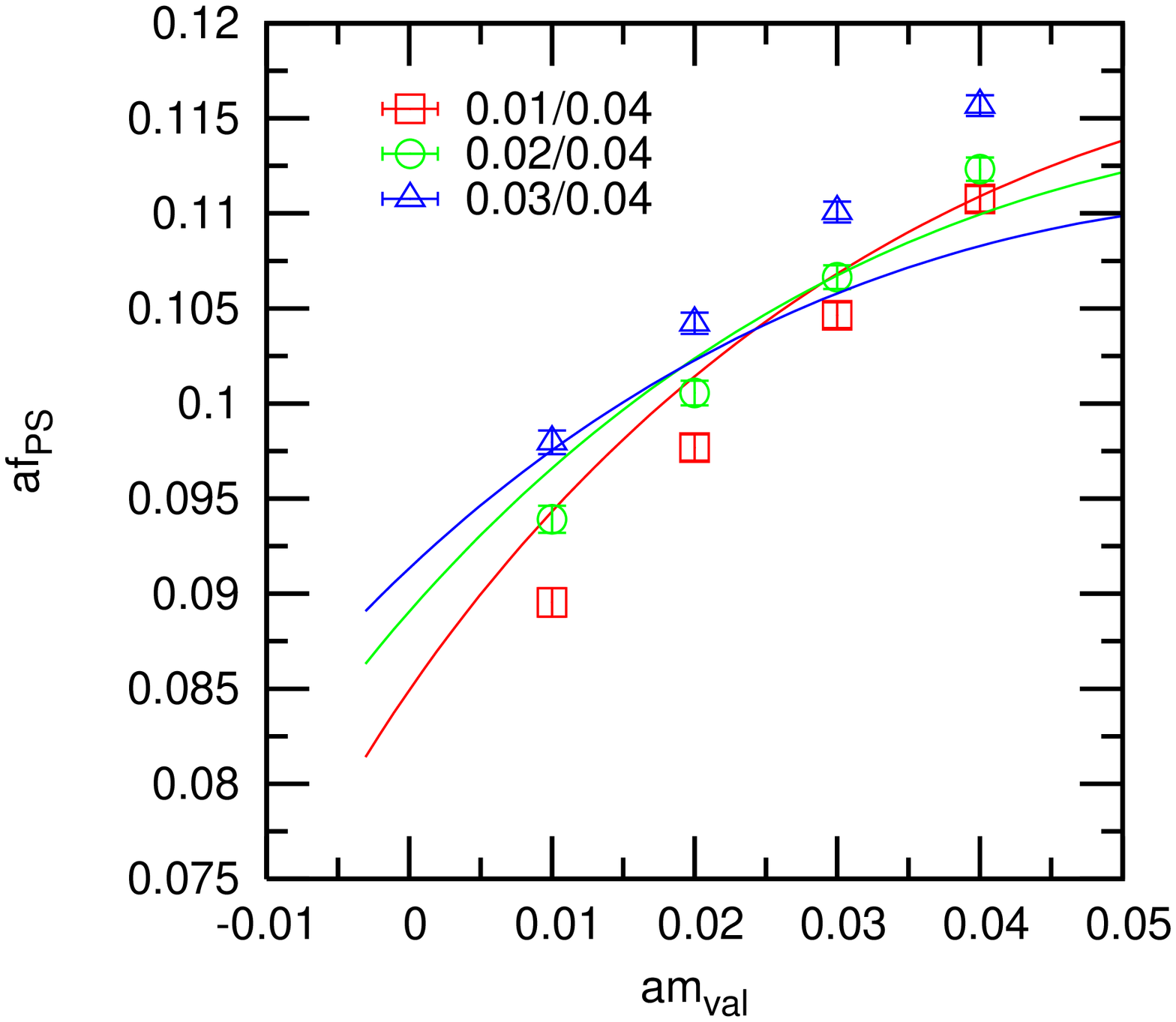}
\end{tabular}
\caption{NLO partially quenched chiral fits on the $16^3\times32$ ensembles. The fits to $M_{PS}^2$ and $f_{PS}$ are carried out \emph{simultaneously}. The data points at $am_{val} = 0.04$ are excluded from the fits.}
\label{fig:sim_fits_n16}
\end{center}
\end{figure}

Neglecting the constant term in Eq.~\ref{eq:mpi_NLO}, the total number of the unknown parameters is 6. Since both $M_{PS}^2$ and $f_{PS}$ depend on $B_0$ and $f$, they should be included in the chiral fits simultaneously. For comparison, we first fit the pseudoscalar masses and decay constants independent of each other with the 0.04 valence point excluded from the fits (Figure~\ref{fig:independent_fits}). The data is described by the NLO chiral formulas amazingly well. However, the values of $B_0$ and $f$ obtained from the fits of $M_{PS}^2$ and $f_{PS}$ differ by a factor of 2. This inconsistency comes from the fact that the lattice data of $f_{PS}$ does not have the curvature predicted by the NLO ChPT. The quark masses we are working with correspond to pion masses of 390 to 630 MeV, which may be outside the chiral region where NLO ChPT is sufficient. It is thus not surprising to see how badly the simultaneous fits to $M_{PS}^2$ and $f_{PS}$ fail, as shown in Figure~\ref{fig:sim_fits_n16}. 
\vspace{-1cm}

\begin{figure}[ht]
\begin{center}
\begin{tabular}{cc}
\includegraphics*[angle=-90,width=0.45\textwidth,clip]{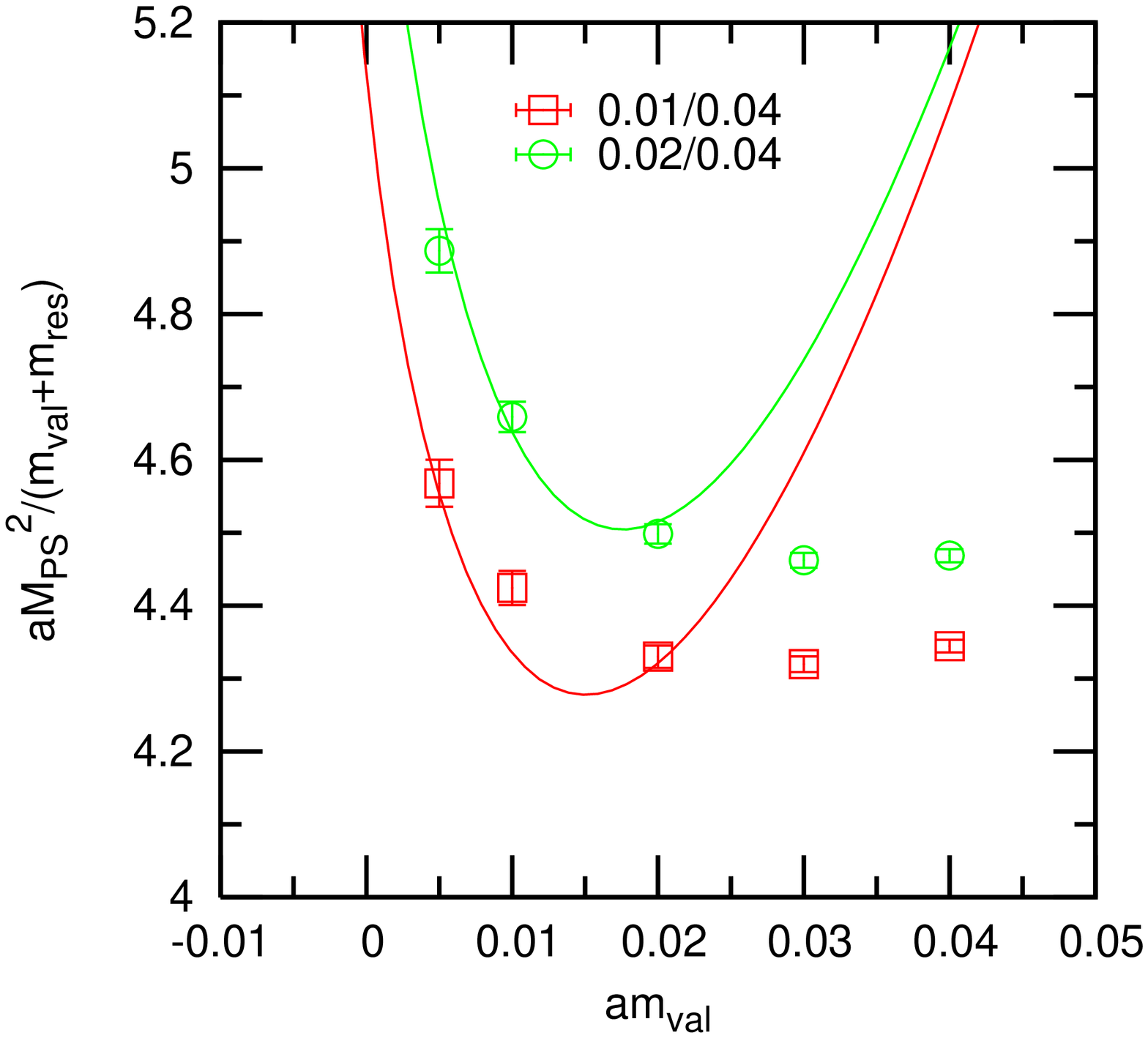} &
\includegraphics*[angle=-90,width=0.45\textwidth,clip]{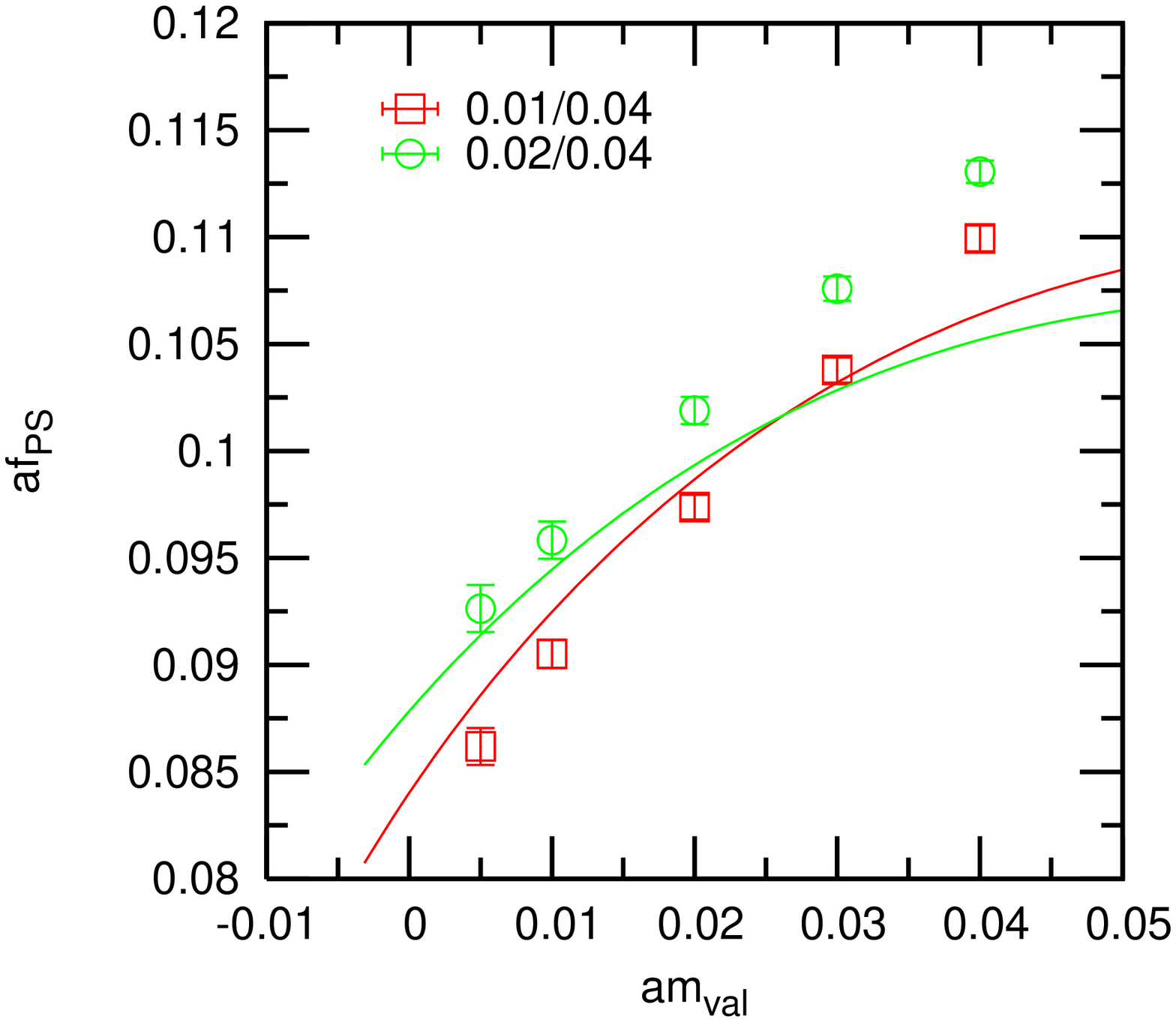}
\end{tabular}
\caption{Partially quenched NLO chiral fits on the $24^3\times64$ ensembles. The fits to $M_{PS}^2$ and $f_{PS}$ are performed \emph{simultaneously} in the quark mass range of 0.005 to 0.02.}
\label{fig:sim_fits_n24}
\end{center}
\end{figure}

The failure to fit our data to the NLO chiral formulas in this mass range motivated us to do partially quenched measurements with a lighter valence mass of 0.005 on the $24^3\times64$ lattices. In Figure~\ref{fig:sim_fits_n24} simultaneous fits for $M_{PS}^2$ and $f_{PS}$ to the 0.005 to 0.02 mass points are plotted. Although the fitted curves do not perfectly represent the data, as indicated by the large $\chi^2$/dof in Table~\ref{tab:pars}, we can see the tendency of  them being more consistent with the data as the quark masses go lighter. The fitted curves miss the heavy mass points badly, which may indicate that the NNLO contributions are significant beyond the mass range in the fits.

\vspace{-0.5cm}
\begin{table}[ht]
\begin{center}
\begin{tabular}{ccccccccc}
\hline
Fit & $\chi^2$/dof & $2B_0a$ & $fa$ & $2L_6-L_4$ & $2L_8-L_5$ & $L_4$ & $L_5$ & $\rho/2B_0$ \\
\hline
I & 9.8(3.6) & 4.12(10) & 0.0555(15) & 0.56(13) & 3.14(32) & -1.05(34) & 4.95(86) & - \\
II & 10.5(3.4) & 3.58(19) & 0.0550(19) & 0.83(16) & 5.5(1.1) & -0.66(49) & 5.04(93) & 9.2(3.7) \\
\hline
\end{tabular}
\caption{Values of the low energy constants determined from the partially quenched NLO chiral fits to $M_{PS}^2$ and $f_{PS}$ without and with the ${\cal O}(a)$ corrections for the $24^3\times64$ ensembles.  The quark masses included in the fits are from 0.005 to 0.02. The values of $L_i$'s and $\rho/2B_0$ have been multiplied by $10^4$. The chiral scale $\mu$ is taken to be 1 GeV. }
\label{tab:pars}
\end{center}
\end{table}
\vspace{-0.2cm}
\subsubsection{${\cal O}(a)$ corrections}
Since the quark masses we have measured on the $16^3\times32$ lattices are too heavy to have consistent chiral fits, in this section we only consider the data for the $24^3\times64$ ensembles. Including ${\cal O}(a)$ effects in the chiral fits only introduces one more parameter $\rho$ in Eq.~\ref{eq:mpi_NLO}. We expect the size of $\rho/2B_0$ to be much smaller than $m_q$ as discussed in Section~\ref{sec:DWF}. Applying such fits to the data does not change the fit quality very much. It only slightly shifts the values of the low energy constants determined from the fits. Table~\ref{tab:pars} compares the fitted parameters from the fits without and with the ${\cal O}(a)$ corrections, denoted as Fit I and Fit II respectively. As expected, the value of $\rho/2B_0$ is about 0.001, which is 1/3 of $\mres$, or 1/8 of the smallest quark mass (0.005+0.003) in the fits. This is consistent with the assumption we make to employ the whole power counting scheme that the ${\cal O}(a)$ chiral symmetry breaking effect is small compared to the mass term for domain wall fermions.

\section{Summary and Future Plans}
To summarize, the finite lattice spacing corrections for domain wall
fermions are found to be small, with our current masses and
statistics. It is possible to apply the continuum chiral formulas
directly in the chiral fits. However, the quark masses in current
simulations are still too heavy to see a perfect consistency between
the predictions of ChPT and the numerical results. To answer the
question of how light the quark masses should be to have reliable
chiral extrapolations, we need to perform simulations at even lighter
quark masses. Our measurements on the $24^3\times64$ ensembles with a
valence quark mass of 0.001 are in progress. The RBC and UKQCD collaborations are also generating a new $24^3\times64$ ensemble with the light sea quark mass of 0.005. These new data points will give us a good opportunity to probe the chiral limit for various quantities. We also plan to study the chiral extrapolations at NNLO with the full formulas calculated by Bijnens \textit{et al.}~\cite{Bijnens:2006jv}.

\section*{Acknowledgments}

We thank Peter Boyle, 
Dong Chen, Mike Clark, Saul Cohen, Calin Cristian, Zhihua Dong, Alan Gara, 
Andrew Jackson, Balint Joo, Chulwoo Jung, Richard Kenway, Changhoan Kim,
Ludmila Levkova, Xiaodong Liao, Guofeng Liu, Robert Mawhinney, Shigemi Ohta, 
Konstantin Petrov, Tilo Wettig and Azusa Yamaguchi for developing with us 
the QCDOC machine and its software. This development and the resulting 
computer equipment used in this calculation were funded by the U.S. DOE 
grant DE-FG02-92ER40699, PPARC JIF grant PPA/J/S/1998/00756 and by RIKEN. 
This work was supported by DOE grant DE-FG02-92ER40699 and we thank RIKEN,
Brookhaven National Laboratory and the U.S. Department of Energy for 
providing the facilities essential for the completion of this work.

\bibliographystyle{apsrev}
\bibliography{mflin}

\end{document}